\documentclass[conference]{IEEEtran}

\usepackage{graphicx}
\usepackage{cite}
\usepackage{amsmath}
\usepackage{color}
\usepackage{subfig}
\usepackage{amssymb}

\title{Adaptive Single-Terminal Fault Location for DC Microgrids}

\author{
\IEEEauthorblockN{Vaibhav Nougain\IEEEauthorrefmark{1}, Sukumar Mishra\IEEEauthorrefmark{2}, Joan-Marc Rodriguez-Bernuz\IEEEauthorrefmark{3}, Adrià Junyent-Ferré\IEEEauthorrefmark{4}, \\
Aditya Shekhar\IEEEauthorrefmark{1}, Aleksandra Lekić\IEEEauthorrefmark{1}}
\vspace{0.05in}
\IEEEauthorblockA{\IEEEauthorrefmark{1}Delft University of Technology, Dept. Electrical Sustainable Energy, Delft, Netherlands \\
Emails: \{V.Nougain, A.Shekhar, A.Lekic\}@tudelft.nl}
\IEEEauthorblockA{\IEEEauthorrefmark{2}Indian Institute of Technology Delhi, Dept. of Electrical Engineering, Delhi, India,  Email: sukumar@ee.iitd.ac.in}
\IEEEauthorblockA{\IEEEauthorrefmark{3}Technical University of Catalonia, Dept. of Electrical Engineering, Barcelona, Spain,  Email: joan.marc.rodriguez@upc.edu}
\IEEEauthorblockA{\IEEEauthorrefmark{4}Imperial College London, Dept. of Electrical and Electronic
 Eng., London, UK  Email: adria.junyent-ferre@imperial.ac.uk}

}

\begin{document}

\maketitle

\begin{abstract}
Identifying faulty lines and their accurate location is key for rapidly restoring distribution systems. This will become a greater challenge as the penetration of power electronics increases, and contingencies are seen across larger areas. This paper proposes a single terminal methodology (i.e., no communication involved) that is robust to variations of key parameters (e.g., sampling frequency, system parameters, etc.) and performs particularly well for low resistance faults that constitute the majority of faults in low voltage DC systems. The proposed method uses local measurements to estimate the current caused by the other terminals affected by the contingency. This mimics the strategy followed by double terminal methods that require communications and decouples the accuracy of the methodology from the fault resistance. The algorithm takes consecutive voltage and current samples, including the estimated current of the other terminal, into the analysis. This mathematical methodology results in a better accuracy than other single-terminal approaches found in the literature. The robustness of the proposed strategy against different fault resistances and locations is demonstrated using MATLAB simulations.
\end{abstract}


\section{Introduction}

\IEEEPARstart {L}{ow} voltage DC (LVDC) will play a vital role in future low-carbon electrical energy systems with solar photovoltaics, electric vehicles, etc. \cite{ref1}. Potentially, LVDC will enable the use of power-denser hardware and bring less stand-by losses, more control flexibility, and higher power quality \cite{ref2, ref3, ref4}. The protection of DC systems requires reliability, security, dependability, and selectivity \cite{ref4} to identify and locate a fault. Therefore, fault identification techniques help reduce the impact of contingencies, whereas location methods allow a timely restoration of
the system. Fault location methods are divided into two categories in the literature: offline and online methods. Offline methods calculate the fault location after the fault has been cleared. In contrast, online methods evaluate the fault distance after the fault has been identified but before it is cleared.

Discussing the offline methods in the literature, the authors in \cite{ref7, ref8, ref9} use a probe power unit (PPU) as an active external injection unit. The authors in \cite{ref7} neglect the damping coefficient in the second-order RLC discharging circuit, taking the damped frequency to be equal to the natural frequency. This results in a loss of accuracy of the fault distance location. Authors in \cite{ref8} use a least squares algorithm to evaluate the damping i.e., attenuation constant. However, the method's accuracy is compromised when the attenuation constant is high. Paper \cite{ref9} uses a modified PPU to calculate a high value for the attenuation constant and, therefore, evaluate the fault distance with higher accuracy compared to \cite{ref7, ref8, ref9, nougain2023fault}. The authors in \cite{ref5} use a fault location module consisting of an inductor, two switches, and two thyristors to achieve an accuracy of 98.4$\%$ for low resistance faults in an LVDC system. However, the method in \cite{ref5} is prone to white Gaussian noise (WGN) in measurement as it uses a differential term (\textit{di/dt}) in the analysis of the fault location.
Authors in \cite{ref6} propose an iterative method to evaluate the fault location for pole-to-ground (PTG) faults without any external modification. However, this method has a dead zone for faults closer to the bus and for pole-to-pole (PTP) faults \cite{ref10}. Additionally, it is desirable to have a robust online method over its offline counterpart.
\begin{table*}[!t]
\renewcommand{\arraystretch}{1.3}
\caption{Fault location methods in literature}
\centering
\label{table_1}
\resizebox{0.85\linewidth}{!}{%
		\begin{tabular}{c c c c}
			\hline\hline \\[-3mm]
			\multicolumn{1}{c}{Fault Location Methods} & \multicolumn{1}{c}{Location Sequence/Differential Term}  & \multicolumn{1}{c}{Other Terminal Considered?} & \multicolumn{1}{c}{Degree of External Modification Required?}
			\\[1.6ex] \hline
			\cite{ref10}, \cite{ref16}  &  Online/No & No & low, current limiting reactor \\
			\hline
			\cite{ref11}  & Online/No & No & high, improved solid-state circuit breaker \\
			\hline
		    \cite{ref12}  & Online/Yes, prone to WGN & No & high, multiple capacitance earthing \\
		    \hline
			\cite{ref13}  & Online/No & Yes, estimates other terminal current & Not required, high sampling frequency\\ 
			\hline
			\cite{ref5}  &  Offline/Yes, prone to WGN & No & high, Fault location module \\
			\hline
			\cite{ref14}-\cite{ref15} & Online/Yes, prone to WGN & Yes, require synchronous data  & Not required \\
			\hline
			\cite{ref6} &  Offline/Yes, prone to WGN & No & Not required \\
			\hline
			\cite{ref7}-\cite{ref9}  &  Offline/No & No & high, Probe power unit\\
			\hline
			\cite{ref17}  &  Online/No & Yes, requires asynchronous data  & high, modified T-source circuit breaker \\
			\hline
			[Proposed method] &  Online/No & Yes, estimates other terminal current   & low, current limiting reactor \\
			 [1.4ex]
			\hline\hline
\end{tabular}%
}
\end{table*}
Regarding online methods, the authors in \cite{ref10} propose using a current limiting reactor (CLR) in single and double terminal fault location methods. The analysis is robust and convincing based on the ratio of transient voltages (ROTV). The authors in \cite{ref11} use an improved solid-state circuit breaker (SSCB) to propose a curve-fitting method. However, the parameter extraction method presents low accuracy when the fault occurs close to the bus \cite{ref17}. Paper \cite{ref12} uses multiple capacitive earthing points as an external modification that can be used to locate faults. The authors in \cite{ref14,ref15} use current and voltage measurements from both ends to formulate a fault location method without any external modification. The method from \cite{ref14} uses a Moore–Penrose pseudo-inverse, whereas the authors in \cite{ref15} use an iterative approach for fault location. However, the performance in \cite{ref14}, \cite{ref15} and \cite{ref12} is affected by WGN. Moreover, papers \cite{ref14, ref15} require synchronized double terminal data for accurate fault location. The authors in \cite{ref16} propose a Newton-Raphson-based fault detection and location method for low-resistance faults and a ground current relay to locate high-resistance faults (up to 25$\Omega$). However, this method assumes an overdamped response of current and voltages in the analysis, even for low-resistance faults. This reduces the accuracy of the fault location strategy under low fault resistance scenarios. The authors in \cite{ref17} use a modified T-source circuit breaker in a method that is robust to WGN and effective to high fault resistance. However, the method requires double terminal asynchronous data to eliminate the influence of fault resistance.

This paper presents an online single-terminal fault location method for flexible (point-to-point or multi-terminal) LVDC systems. The novelty of this method lies in improving the fault location accuracy for low resistance faults up to 1 $\Omega$, which constitute the majority of faults in LVDC systems (usually with low resistive earthing \cite{ref12, ref37}). Low voltage systems have a high possibility of experiencing low resistance pole-to-ground faults. Likewise, short circuit pole-to-pole faults are less likely to occur, and high resistance pole-to-ground faults are rare \cite{ref7, ref11}.

The proposed approach is based on estimating the current of other terminals that are being directly affected during a fault using only local measurements. Including the estimated current in the single terminal iterative fault location method improves fault location accuracy. However, as the fault resistance increases, the fault current contribution in the analysis reduces. As a result, the merit of using estimated current reduces for high fault resistances. The proposed method shows similar accuracy to conventional single terminal methods \cite{ref7, ref8, ref9, ref10} for higher fault resistances up to 5 $\Omega$. Therefore, the fault location method is accurate and applicable for a conventional LVDC system with a fault resistance ranging from 0-5 $\Omega$. Briefly, the proposed location method has the following key attributes:
\begin{itemize}
  \item Single-terminal. The proposed method does not require information from other terminals to locate the fault. This reduces its cost and complexity. However, single-terminal methods are dependent on fault resistance. The proposed work estimates other terminals' current to mimic the double terminal methods' methodology. This eliminates the dependence of fault location accuracy on fault resistance, improving the accuracy of fault location.
  \item No differential term. The method is robust against WGN in measurements as it replaces the differential current term with the voltage across CLRs since the use of differential terms (\textit{di/dt} and \textit{dv/dt}) in the mathematical analysis make these approaches prone to the loss of accuracy in the presence of WGN in measurements. 
  \item Minimum external modification. To optimise the cost of the LVDC systems, it is important to do minimum changes to the existing systems for the purpose of control, operation and protection. The proposed method requires a low degree of external modification in the form of CLRs \cite{ref10,ref16}, which is widely suggested and used in literature for the protection of DC systems. This reduces the cost and complexity of the fault location method compared to other strategies such as \cite{ref11} (improved solid-state circuit breakers), \cite{ref12} (multiple capacitive earthing points), \cite{ref5} (fault location module), \cite{ref17} (modified T-source circuit breaker).
\end{itemize}
\begin{figure*}
	\centering
	\includegraphics[width=0.9\linewidth,  height= 8.5cm]{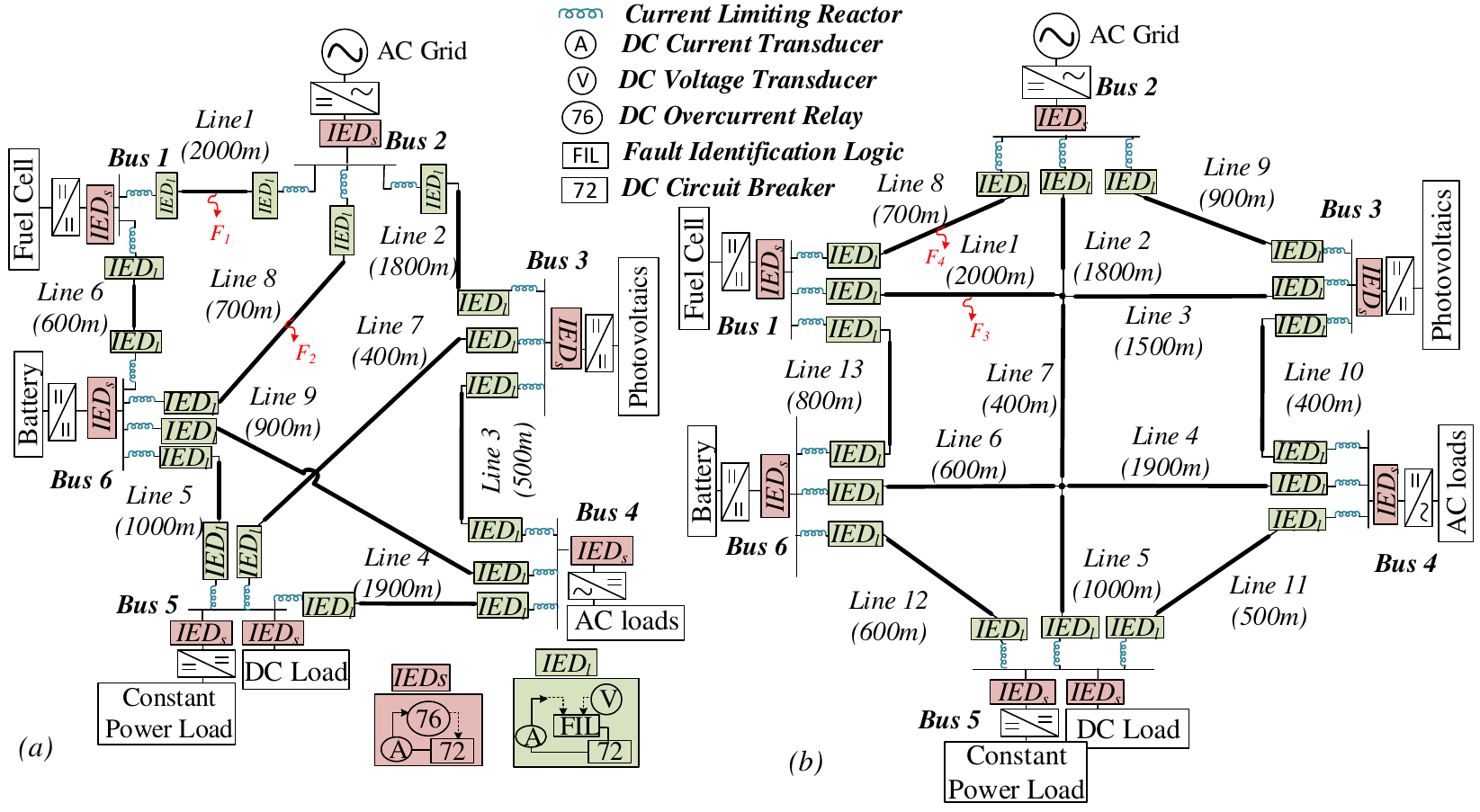}	\caption{LVDC Test System formed by (a) point-to-point, (b) multi-terminal configuration}
	\label{fig:sys_model}
 \vspace*{-5mm}
\end{figure*}

The rest of the paper is organized as follows. Section II describes the test system. Section III presents the fault location algorithm using the estimation of the current experienced by other terminals and the related validation in MATLAB/Simulink. Finally, section IV brings the conclusions of the work.
\section{Test System Configuration}
The fault location algorithm is developed to be used either in a point-to-point or multi-terminal distribution system. Therefore, two systems with identical elements but different configurations (i.e. one as a multi-terminal and the other as a point-to-point) are considered in Fig. \ref{fig:sys_model}.
These systems operate at $\pm$380V and are formed by 6 buses. The LVDC system consists of a fuel cell on Bus 1, an AC grid on Bus 2, a photovoltaic system on Bus 3, and an energy storage system on Bus 6. Buses 4-5 have AC loads, DC loads, and constant power loads \cite{ref18}. Benchmark controllers used in conventional LVDC system applications have been implemented to validate the analysis \cite{ref22}.

\begin{figure}
	\centering
	\includegraphics[width=0.9\columnwidth,  height= 5.5cm]{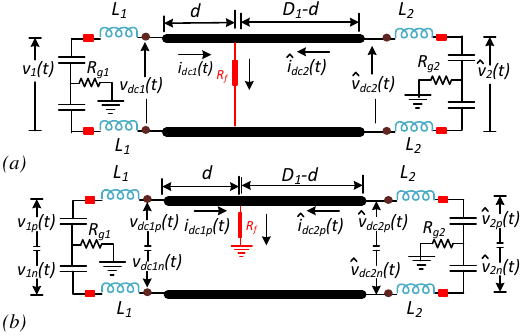}
	\caption{Simplified network for point-to-point configuration under (a) PTP fault, (b) P-PTG fault}
 \textbf{\captionsetup{justification = raggedright,
singlelinecheck = false,font=footnotesize}}
	\label{fig:simple_sys_model}
  \vspace*{-1cm}
\end{figure}

\begin{figure}
	\centering
	\includegraphics[width=0.75\columnwidth,  height= 4.8cm]{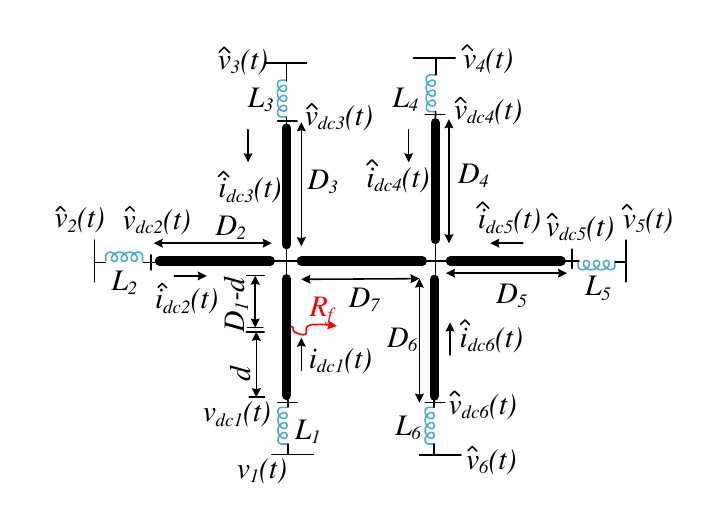}
	\caption{Simplified multi-terminal test system}
 \captionsetup{justification = raggedright,
singlelinecheck = false,font=footnotesize}
	\label{fig:simple_mtdc}
\end{figure}

The system needs the addition of CLRs to limit the rate of change of the current during faults. For the point-to-point configuration, CLRs are present at both cable terminals. However, for a multi-terminal configuration, CLRs may not be present at both cable terminals (see Fig. \ref{fig:sys_model}(b)). As a result, the fault location method is independently analyzed for point-to-point and multi-terminal systems. For the point-to-point system, the fault contribution from other buses is neglected since the converter´s controllers' slow bandwidth and the path´s high inductance would limit the current from other buses \cite{ref14,ref15}. This permits simplifying the faulty point-to-point network as shown in Fig. \ref{fig:simple_sys_model}. Similarly, for the multi-terminal system, the contribution from other buses is neglected, but the contribution from other nodes is considered (see Fig. \ref{fig:simple_mtdc}). This makes the analysis of fault location slightly tedious for the multi-terminal system compared to the point-to-point system. Fig. \ref{fig:simple_mtdc} is used as the effective and simplified single-line network for a multi-terminal system.

A bipolar line configuration with a frequency-dependent transmission model (FDTL) of an underground cable (UGC) is considered. The LVDC system consists of different cables modeled as single-core cables with copper conductor, XLPE insulation (2.8 mm thick), and PVC sheathing (2.5 mm thick). The grounding capacitance of the UGC is 0.5$\mu$F/km while the equivalent DC capacitor of the converter is around 10$^{3}$-10$^{4}$ $\mu$F. Hence, the fault contribution from the grounding capacitance of cables and lines can be ignored for short to medium cables. As a result, the $R-L$ representation of UGC is used for the fault location analysis. The grounding scheme considered is TN-S, which provides mid-point grounding at each converter terminal. The fault contingencies considered in the test system are of different types, i.e., PTP, P-PTG, and N-PTG. The sources and loads in the setup are protected using an intelligent electronic device on the source side (\textit{IED$_{s}$}) which is a combination of DC current transducer and DC over-current relays and DC circuit breakers as shown in Fig. 1. The cables in the system are protected using intelligent electronic devices of load side (\textit{IED$_{l}$}) \cite{ref4,ref19}.  The fault current used in the analysis is supplied by the discharging DC bus capacitance, not the inverters and converters upon fault inception. The converters should trip as soon as overcurrent hits. This is indicated in Fig. \ref{fig:sys_model} using DC circuit breakers in \textit{IED$_{s}$} and \textit{IED$_{l}$}.

\section{Fault Location Algorithm}
\subsection{Estimation of other terminal current, $\hat{i}_{dc2}(t)$}
Fig. \ref{fig:sys_model}(a) shows the point-to-point system with a fault between bus 1 and bus 2 (shown as \textit{F$_{1}$} in Fig. \ref{fig:sys_model}(a)). Fig. \ref{fig:simple_sys_model}(a) shows the simplified network under a PTP fault, whereas Fig. \ref{fig:simple_sys_model}(b) shows the simplified network under a P-PTG fault. Since the method is single terminal, $\hat{v}_{dc2}(t)$ and $\hat{i}_{dc2}(t)$ are unknowns \cite{ref28}. The analysis assumes $v_{1}(t) \approx \hat{v}_{2}(t)$ as voltage collapses slowly due to the energy dynamics of the DC bus as a result of its equivalent capacitance. Applying KVL in Fig. \ref{fig:sys_model}(a), following equations are obtained \cite{ref27}:
\begin{equation}
v_{dc1}(t)=2rdi_{dc1}(t)+2ld\frac{di_{dc1}(t)}{dt}+R_{f}[i_{dc1}(t)+\hat{i}_{dc2}(t)],
\label{eq:1} \normalsize
\end{equation}
\vspace{-1cm}
\begin{multline}
\small
\hat{v}_{dc2}(t)=2r(D_1-d)\hat{i}_{dc2}(t)+2l(D_1-d)\frac{d\hat{i}_{dc2}(t)}{dt}\\+R_{f}[i_{dc1}(t)+\hat{i}_{dc2}(t)],
\end{multline}
where \textit{v$_{dc1}(t)$} and \textit{$\hat{v}_{dc2}(t)$} are the terminal voltages at each side of the cable, \textit{i$_{dc1}(t)$} and \textit{$\hat{i}_{dc2}(t)$} are the currents fed into the fault by the terminals on each side. \textit{r} and \textit{l} are the unit resistance and inductance of UGC. $L_{n}$ is defined as the value of the CLR, and $R_{f}$ is defined as the fault resistance. $D_{1}$ is the cable length whereas $d$ is the fault distance from bus terminal. Considering the voltage drop across the CLR, the current derivative terms can be obtained, avoiding substitution errors due to differential calculations \cite{ref2}. This improves the algorithm's performance in the presence of WGN. These terms are given as $\frac{di_{dc1}(t)}{dt}=\frac{v_{1}(t)-v_{dc1}(t)}{2L_{1}}$ and $\frac{d\hat{i}_{dc2}(t)}{dt}=\frac{\hat{v}_{2}(t)-\hat{v}_{dc2}(t)}{2L_{2}}$. Dividing (1) with $v_{1}(t)$ and (2) with $\hat{v}_{2}(t)$, (3)-(4) are obtained, where the ratio of transient voltages is defined as, $\gamma_{1}(t)=\frac{v_{dc1}(t)}{v_{1}(t)}$ and $\hat{\gamma}_{2}(t)$=$\frac{\hat{v}_{dc2}(t)}{\hat{v}_{2}(t)}$.
\begin{equation}
\gamma_{1}(t)=\frac{L_{1}}{L_{1}+ld}\Bigg[\frac{2rdi_{dc1}(t)}{v_{1}(t)}+\frac{ld}{L_{1}}+R_{f}\frac{i_{dc1}(t)+\hat{i}_{dc2}(t)}{v_{1}(t)}\Bigg]
\end{equation}
\begin{multline}
\hat{\gamma}_{2}(t)=\frac{L_{2}}{L_{2}+l(D_1-d)}\Bigg[\frac{2r(D_1-d)\hat{i}_{dc2}(t)}{\hat{v}_{2}(t)}+\frac{l(D_1-d)}{L_{2}}\\+R_{f}\frac{i_{dc1}(t)+\hat{i}_{dc2}(t)}{\hat{v}_{2}(t)}\Bigg]
\end{multline}
Considering (3) and discussing $\frac{2rdi_{dc1}(t)}{v_{1}(t)}$, generally $2rd<1$ for LVDC systems and $\frac{i_{dc1}(t)}{v_{1}(t)}$ is at its minima right after the fault inception. During the fault transient, i$_{dc1}(t)$ rises and $v_{1}(t)$ drops which cumulatively increase the value of $\frac{i_{dc1}(t)}{v_{1}(t)}$. 

Similarly, $\frac{i_{dc1}(t)+\hat{i}_{dc2}(t)}{v_{1}(t)}$ in $R_{f}\frac{i_{dc1}(t)+\hat{i}_{dc2}(t)}{v_{1}(t)}$ is at its minima right after the fault inception. During the fault transient, $i_{dc1}(t)$ and $\hat{i}_{dc2}(t)$ rise and $v_{1}(t)$ drops which cumulatively increase the value of $\frac{i_{dc1}(t)+\hat{i}_{dc2}(t)}{v_{1}(t)}$. High resistance faults (HRFs) can increase the value of the time-varying term, but the effect is not dominant enough to compensate low $\frac{i_{dc1}(t)+\hat{i}_{dc2}(t)}{v_{1}(t)}$ right after the fault. Further, as the maximum value of a fault resistance under the analysis is limited to $5 \Omega$ (considering an LVDC system), the effect of this term is fairly limited. Additionally, the total sum of time-varying terms is $R_{f}\frac{i_{dc1}(t)+\hat{i}_{dc2}(t)}{v_{1}(t)}+\frac{2rdi_{dc1}(t)}{v_{1}(t)}\ll\frac{ld}{L_{1}}$ right after a fault. This results from a low CLR ($L_{1}$) value used in LVDC systems.
As a result, the time-varying terms in (3)-(4) are neglected immediately after the inception of a fault, as they increase gradually (see Fig. \ref{fig:simple_mtdc}). Therefore, for $t=0^+$ in (3)-(4), time-varying terms are neglected to give,
$\gamma_{1}(0^+)=\frac{ld}{L_{1}+ld}$ and $\hat{\gamma}_{2}(0^+)=\frac{l(D_1-d)}{L_{2}+l(D_1-d)}$.
 \begin{figure}[h]
 \hspace{1cm}
{\includegraphics[width=0.6\linewidth, height=4cm]{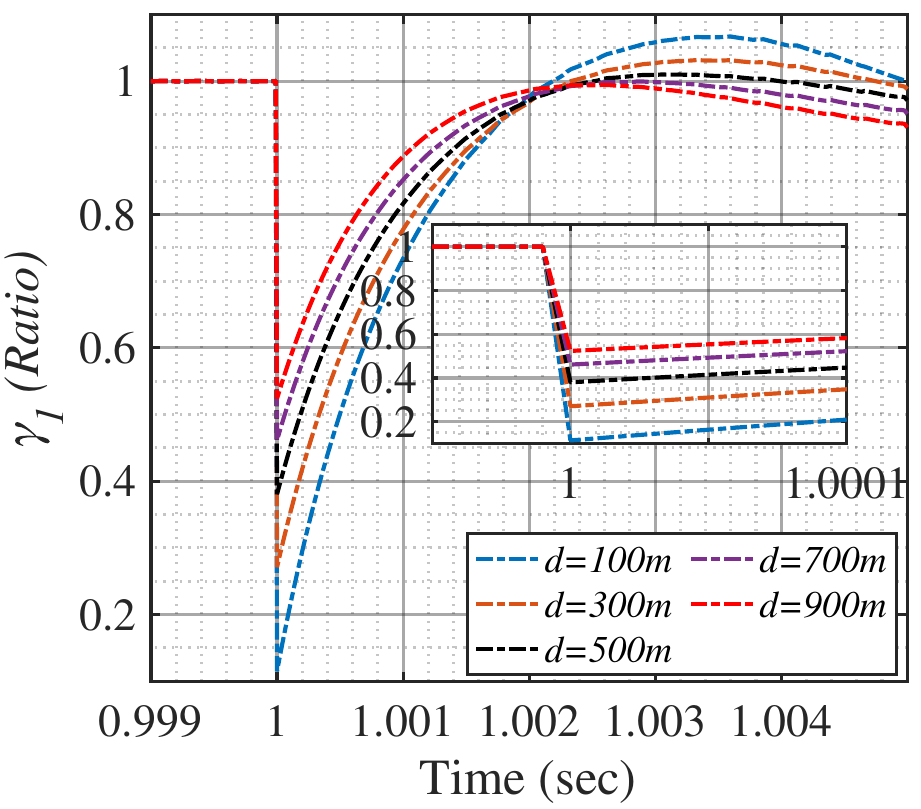}}
\captionsetup{justification = raggedright,
singlelinecheck = false,font=footnotesize}
\caption{Variation of ratio of transient voltages, \textit{$\gamma_{1}$} vs time for different fault distances.}
\label{fig:variation_time_error}
\end{figure} 
A similar analysis can be carried out for a PTG fault using the voltage and current measurements of the faulty pole. By subtracting (3) and (4) and using (5) and (6), the estimated value of other terminal currents for $t=0^+$ is defined as (7a) for a PTP fault and (7b) for a P-PTG fault.
\begin{figure*}[h]
\vspace{-1cm}
\centering
\begin{minipage}{\linewidth}
\centering
\subfloat[]{\includegraphics[width=0.3\linewidth]{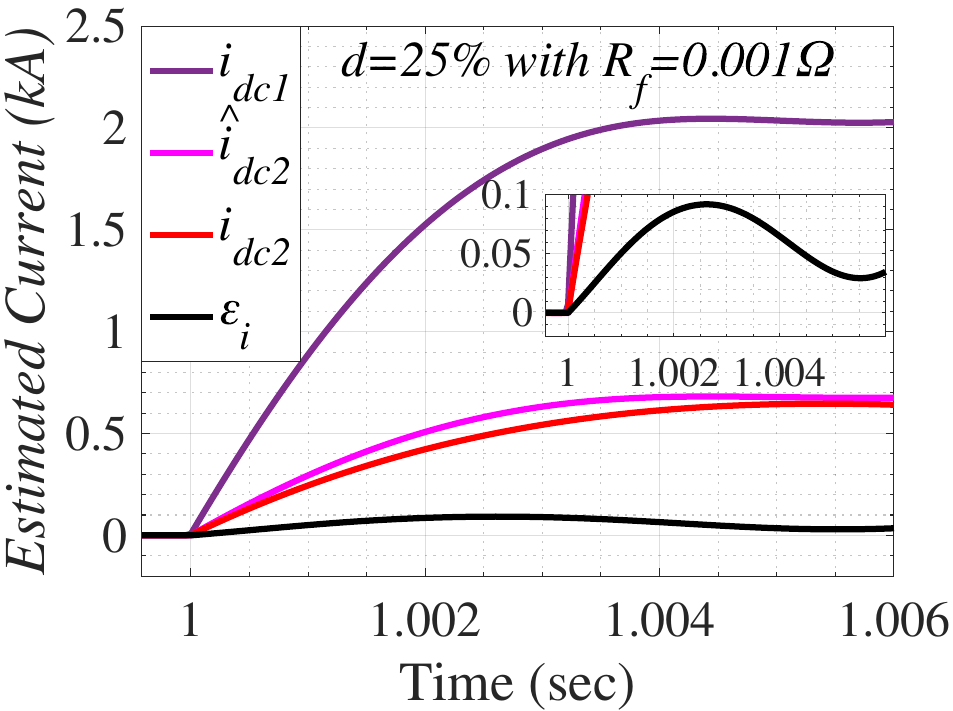}}
\subfloat[] {\includegraphics[width=0.3\linewidth]{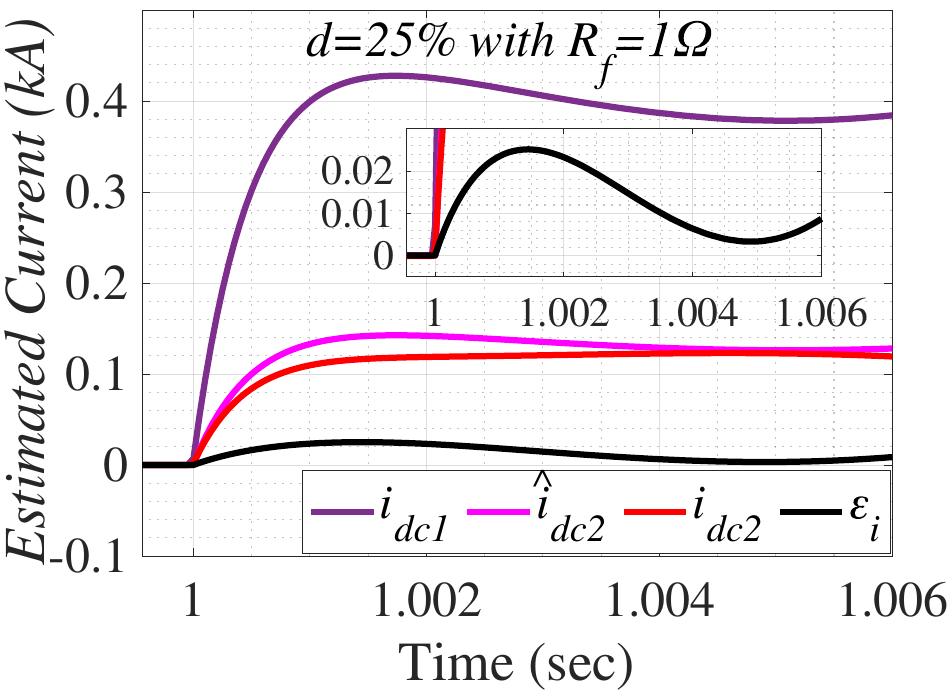}}
\subfloat[] {\includegraphics[width=0.31\linewidth]{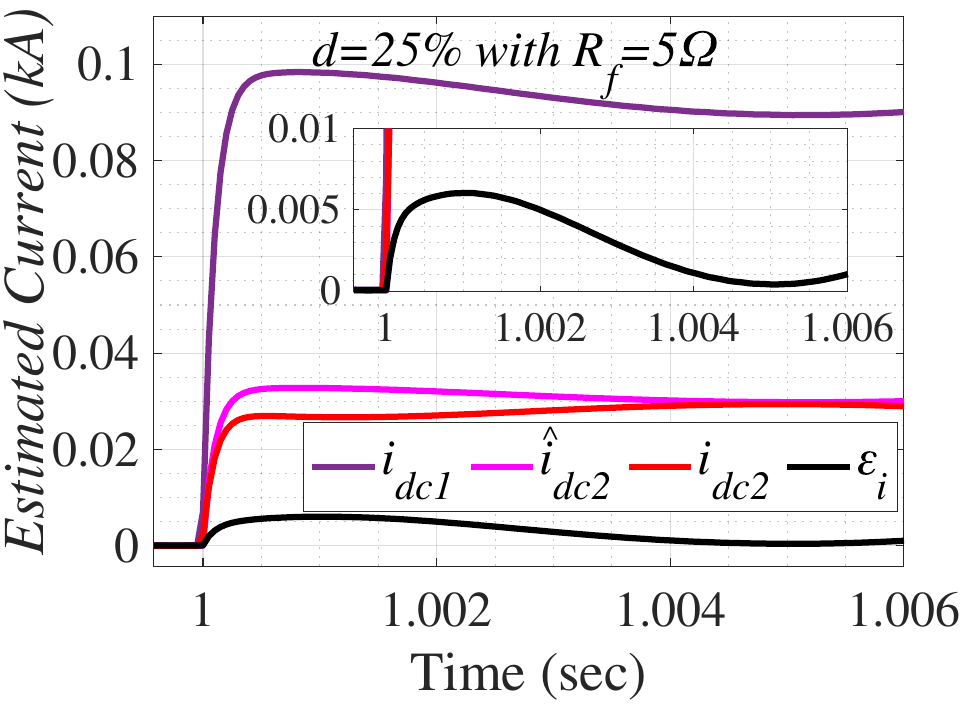}}
\vspace{-1em}
\end{minipage}
\begin{minipage}{\linewidth}
\centering
\subfloat[]{\includegraphics[width=0.3\linewidth]{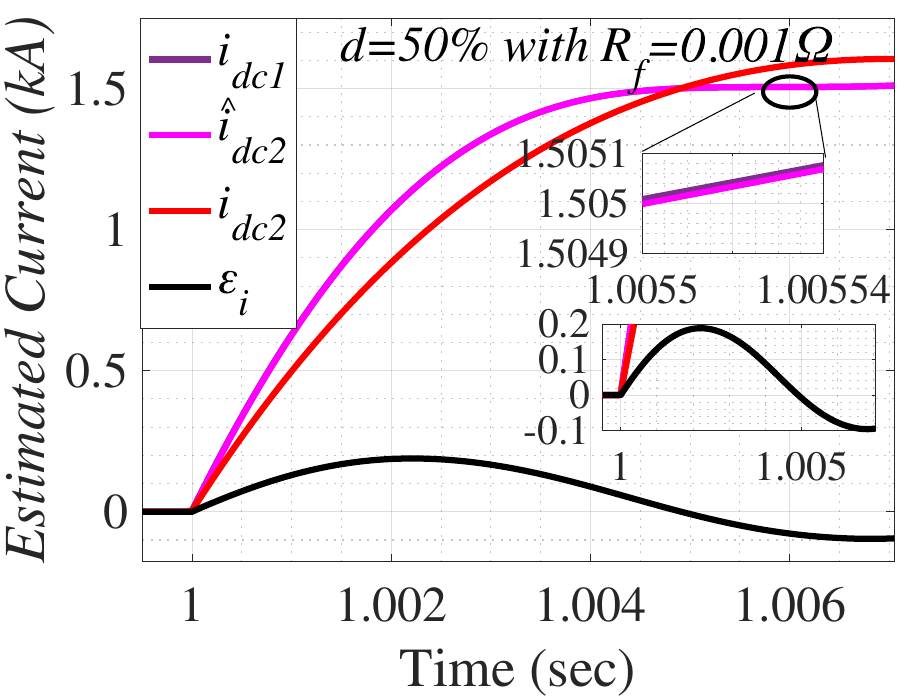}}
\subfloat[] {\includegraphics[width=0.31\linewidth]{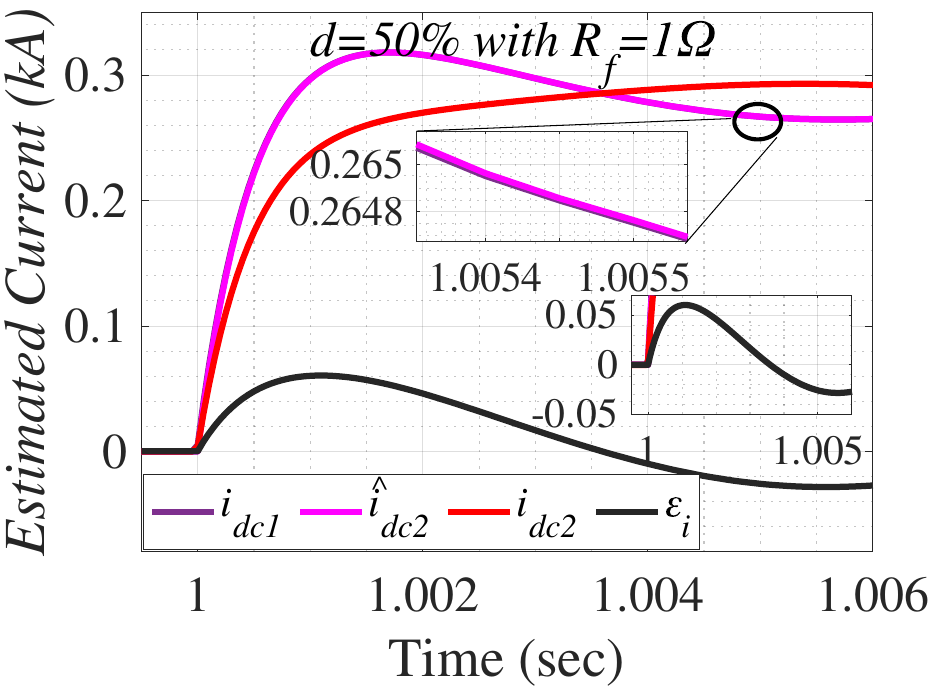}}
\subfloat[] {\includegraphics[width=0.31\linewidth]{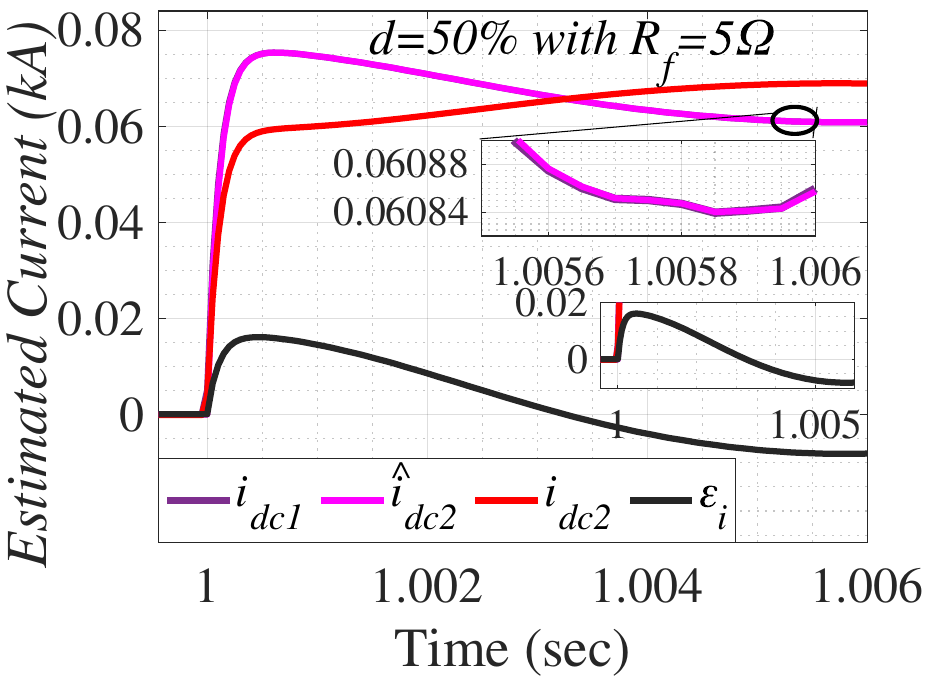}}
\vspace{-0.5em}
\end{minipage}

\captionsetup{justification = raggedright,
singlelinecheck = false,font=footnotesize}
\caption{Estimated terminal current (kA) vs time (sec) for a P-PTG fault at $t=1 \text{ s}$ at 25$\%$ and 50$\%$ with $R_{f}=1 \text{ m}\Omega$, $R_{f}=1\Omega$ and $R_{f}=5\Omega$. }
\label{fig:current_estimated}
\end{figure*} 
 \begin{subequations}
  \begin{eqnarray}
    & \hat{i}_{dc2}(t)=\frac{d}{D_1-d}i_{dc1}(t) : \textit{PTP} \\
    & \hat{i}_{dc2}(t)=\frac{rd+R_{g1}}{r(D_1-d)+R_{g2}}i_{dc1}(t) : \textit{P-PTG}
  \end{eqnarray}
\end{subequations}
The analysis is valid for a point-to-point configuration, but multiple node currents are involved in the analysis for a multi-terminal configuration. As a result, the estimation gets complex.
Fig. \ref{fig:sys_model}(b) shows a multi-terminal LVDC system, whereas Fig. \ref{fig:simple_mtdc} shows its simplified configuration. The objective of the analysis is to find the relation between \textit{$\hat{i}_{dck}(t)|_{k=2}^{6}$} and $i_{dc1}(t)$. The application of KVL considering buses 1-6, simplifying the ratio of transient voltages, gives the relationships between $\hat{i}_{dck}(t)|_{k=2}^{6}$ and $i_{dc1}(t)$. The identities $\hat{i}_{dc3}(t)=\frac{D_{2}}{D_{3}}\hat{i}_{dc2}(t)$, $\hat{i}_{dc4}(t)=\frac{D_{2}}{D_{4}+D_{7}}\hat{i}_{dc2}(t)$, $\hat{i}_{dc5}(t)=\frac{D_{2}}{D_{5}+D_{7}}\hat{i}_{dc2}(t)$ and $\hat{i}_{dc6}(t)=\frac{D_{2}}{D_{6}+D_{7}}\hat{i}_{dc2}(t)$ are used to define $\hat{i}_{dc2}(t)$ in terms of $i_{dc1}(t)$ for $t=0^+$ as in (6).
\begin{equation}
{\small
\hat{i}_{dc2}(t)=\frac{d}{D_{2}+(D_{1}-d)\Bigg[1 + \frac{D_{2}}{D_{3}}+\sum\limits_{k=4}^{6}\frac{D_{2}}{D_{k}+D_{7}}\Bigg]}i_{dc1}(t)}
\end{equation}
For a P-PTG fault ($F_{1}$ in Fig. \ref{fig:sys_model}(a)), Fig. \ref{fig:variation_time_error} shows the time variation of error ($\varepsilon_{i}$) (defined as the difference between estimated other terminal current ($\hat{i}_{dc2}$) and actual current ($i_{dc2}$)). Fig. \ref{fig:current_estimated}(a)-(c) show $\varepsilon_{i}$, $i_{dc2}$, $i_{dc1}$ and  $\hat{i}_{dc2}$ for a fault distance of $25 \%$ with a fault resistance of 1m$\Omega$ (Fig. 5(a)), 1$\Omega$ (Fig. \ref{fig:current_estimated}(b)) and 5$\Omega$ (Fig. \ref{fig:current_estimated}(c)). Similarly, Fig. \ref{fig:current_estimated}(d)-(f) show $\varepsilon_{i}$, $i_{dc2}$, $i_{dc1}$ and  $\hat{i}_{dc2}$ for a fault distance of $50 \%$ with fault resistance of 1 m$\Omega$ (Fig. \ref{fig:current_estimated}(d)), 1$\Omega$ (Fig. \ref{fig:current_estimated}(e)) and 5$\Omega$ (Fig. \ref{fig:current_estimated}(f)). For a fault distance at 50$\%$, $\hat{i}_{dc2}=i_{dc1}$ and the plots in Fig. 5(b),(e),(f) coincide. As the fault resistance increases, fault current reduces, and the merit of using estimated current in the calculation is reduced. Therefore, even if the defined estimated current closely resembles the actual measured current for fault resistance up to 5 $\Omega$, the proposed method shows the merit of better fault location accuracy for fault resistance up to around 1 $\Omega$.
\subsection{Consecutive sample manipulation for fault location}
Due to space constraints, the evaluation of fault location is explained for a point-to-point PTP fault in line with total length, $D_1=2 \text{ km}$. Including the above conclusion in equation\eqref{eq:1}, we can redefine \textit{v$_{dc1}(t)$} as:
\begin{equation}
v_{dc1}(t)=d\bigg(ri_{dc1}(t)+l\frac{di_{dc1}(t)}{dt}\bigg)+R_{f}\frac{D_1}{D_1-d}i_{dc1}(t).
\end{equation}
Rearranging for the calculated fault location, \textit{d}, we can write the following quadratic polynomial: 
\begin{equation}
\begin{array}{l}
d^2\Bigg[r+\frac{l}{L_{m}}\frac{u_{1}(t_{1})}{i_{1}(t_{1})}\Bigg]
-d\Bigg[\frac{v_{dc_{1}}(t_{1})}{i_{1}(t_{1})}+rD_1+\frac{lD_1}{L_{m}}\frac{u_{1}(t_{1})}{i_{1}(t_{1})}\Bigg] \\
+D_1\frac{v_{dc_{1}}(t_{1})}{i_{1}(t_{1})}=D_1R_{f}. 
\end{array}
 \label{eq:8}
\end{equation}
Evaluating equation \eqref{eq:8} at $t=t_{2}$ can give us another equation with the constant $D_1R_{f}$ on the RHS. Subtracting equation\eqref{eq:8} and the equation at $t=t_{2}$ would give us equation \eqref{eq:9}.
\begin{equation}
\small
d^2\frac{l}{L_{m}}\alpha(t_{1},t_{2})
-d\bigg[\beta(t_{1},t_{2})+\frac{lD_1}{L_{m}}\alpha(t_{1},t_{2})\bigg]+D_1\beta(t_{1},t_{2})=0 \label{eq:9}
\end{equation}
where $\alpha(t_{1},t_{2})=[u_{1}(t_{1})i_{1}(t_{2})-u_{1}(t_{2})i_{1}(t_{1})]$ and $\beta(t_{1},t_{2})=[v_{dc1}(t_{1})i_{1}(t_{2})-v_{dc1}(t_{2})i_{1}(t_{1})]$. Continuous subsequent current and voltage measurements~\cite{ref28} are used to solve the quadratic expression.
\begin{figure}
\vspace{-1cm}
\centering
\subfloat[]{\includegraphics[width=0.5\columnwidth]{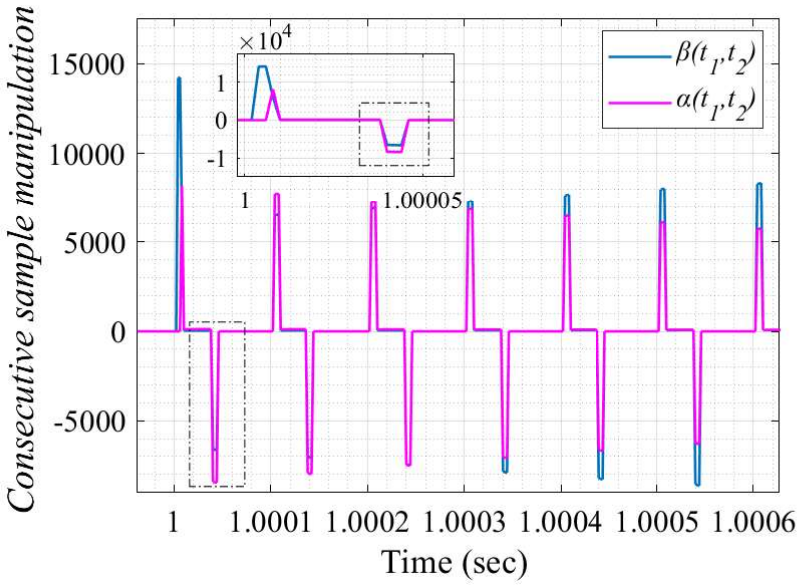}}
\subfloat[]{\includegraphics[width=0.52\columnwidth]{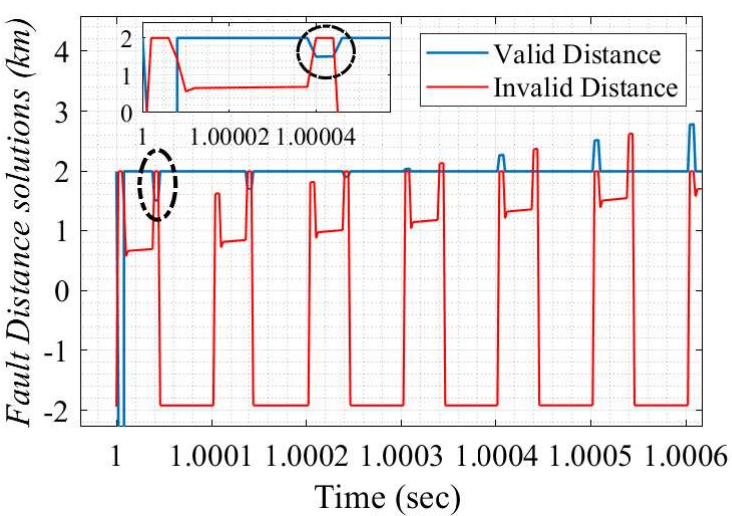}}
\vspace{-0.5em}
\captionsetup{justification = raggedright,
singlelinecheck = false,font=footnotesize}
\caption{(a) $\alpha(t_1,t_2)$ and $\beta(t_1,t_2)$ vs time (sec), (b) Fault distance solutions (km) vs Time (sec) for a line of total length 2 km. }
\label{fig:distance_time}
\end{figure} 
\begin{figure*}[h]
\vspace{-0.4cm}
 \centering
\begin{minipage}{1\linewidth}
\centering
\subfloat[]{\includegraphics[width=0.262\linewidth]{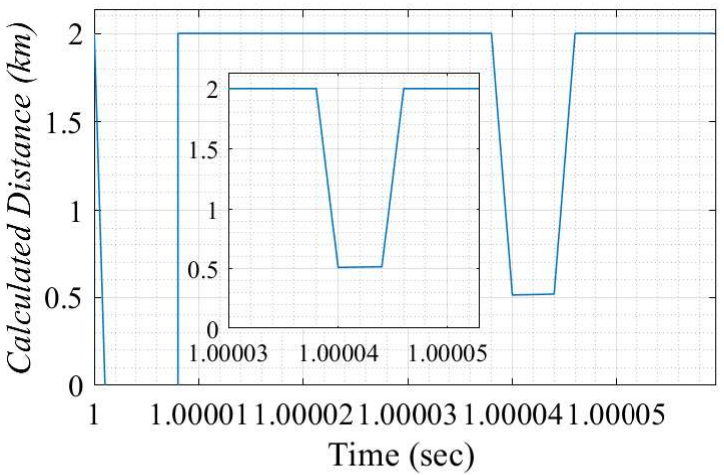}}
\subfloat[] {\includegraphics[width=0.248\linewidth]{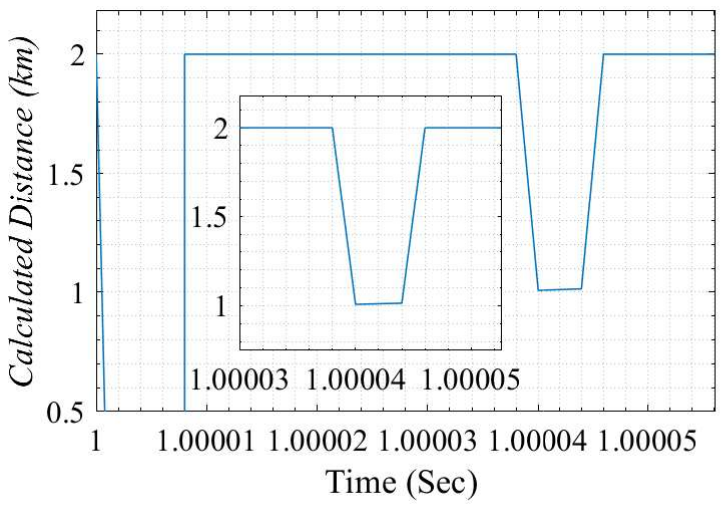}}
\subfloat[] {\includegraphics[width=0.258\linewidth]{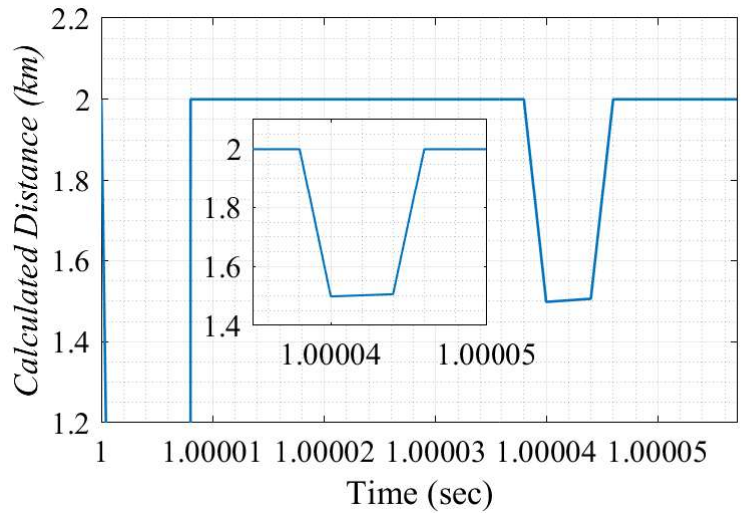}}

\end{minipage}
\captionsetup{justification = raggedright,
singlelinecheck = false,font=footnotesize}
\caption{Calculated fault distance (km) vs time (sec) for a fault at a distance of (a) $d=0.5 \text{ km}$, (b) $d=1 \text{ km}$, (c) $d=1.5\text{ km}$.}
\label{fig:fault_distance}
\vspace*{-5mm}
\end{figure*} 
 Fig. \ref{fig:distance_time}(a) shows the plots for $\alpha(t_{1},t_{2})$ and $\beta(t_{1},t_{2})$ with a window of 3 samples i.e., $t_2=t_1+3\Delta T_s$. A window with fewer samples gives a triangular distance behavior (instead of a trapezoidal behavior) solution, while a window of abruptly high samples, such as 20 samples, gives an erroneous fault distance solution. A window of 3-10 samples is recommended to implement consecutive sample manipulation. The fault information is packed within the black dotted rectangle in Fig. \ref{fig:distance_time}(a) and the black dotted ellipse in Fig. \ref{fig:distance_time}(b). The quadratic polynomial gives two solutions for the fault distance. One of the solutions is negative for the most part and is defined as the invalid distance (shown with red line) in Fig. \ref{fig:distance_time}(b). The valid distance (shown with a blue line) in Fig. \ref{fig:distance_time}(b) gives valuable information on accurate fault distance for nearly 4$\mu$s. Fig. \ref{fig:fault_distance} shows the presence of the DC fault at different locations i.e., $d=0.5 \text{ km}$, $d=1 \text{ km}$ and $d=1.5 \text{ km}$. The algorithm gives accurate fault location for faults at different locations, as evident from Fig. \ref{fig:fault_distance}, where the second transient of the valid solution gives the accurate fault distance. In each case, the fault distance lingers around its true value for nearly 4$\mu$s. This time can be increased with an increase in the window size of consecutive sample manipulation. However, increasing it to a very high value may jeopardize the method's accuracy. The future scope of the work includes generalizing the algorithm for a low-voltage system irrespective of its configuration. Further, the proposed method can be validated in the real-time digital simulator (RTDS).
 \vspace{-0.05cm}
\section{Conclusion} 
The proposed method estimates the adjacent terminal current using local current measurement and mimics double terminal fault location analysis methods. This eliminates the dependence of accuracy on the resistance of the fault. The simple algorithm accurately determines the distance using a consecutive sample manipulation. The proposed methodology presents highly accurate performance for low resistance faults, but as the fault resistance increases, the fault current contribution reduces, and the performance of the approach decreases. Nevertheless, the accuracy for high fault resistances is still comparable to other single terminal methods discussed
in the literature.
\section*{Acknowledgement}
This project has received funding from the European Union’s HORIZON-WIDERA-2021-ACCESS-03 project SUNRISE under grant agreement No. 101079200.
\vspace{-0.05cm}
\bibliographystyle{IEEEtran}
\bibliography{biblo}

\end{document}